 \documentclass{pasj00}

 \SetRunningHead{H.~Imai \etal}
{SiO \unboldmath$v=2$ and $v=$3 $J=1\rightarrow 0$ Maser Features around Evolved Stars}
 \Received{2012/07/31}
 \Accepted{2012/09/16}
 
 \def \kms{~km~s$^{-1}$}
 \def \etal{~et~al.}
 \def \h2o{H$_{2}$O}

\title{Pilot VLBI Survey of SiO $v=$3 $J=1\rightarrow 0$ Maser Emission around Evolved Stars}

\author{Hiroshi  \textsc{Imai}\altaffilmark{1,2}, Miyako \textsc{Oyadomari}\altaffilmark{1}, 
Sze Ning \textsc{Chong}\altaffilmark{1}, Akiharu  \textsc{Nakagawa}\altaffilmark{1}, \\
Tomoharu \textsc{Kurayama}\altaffilmark{3}, Jun-ichi  \textsc{Nakashima}\altaffilmark{4}, 
Naoko \textsc{Matsumoto}\altaffilmark{5}, Takumi  \textsc{Nagayama}\altaffilmark{5}, \\
Tomoaki  \textsc{Oyama}\altaffilmark{5}, Shota \textsc{Mizuno}\altaffilmark{5}, 
Shuji  \textsc{Deguchi}\altaffilmark{6}, and Se-Hyung \textsc{Cho}\altaffilmark{7}}

\altaffiltext{1}{Graduate School of Science and Engineering, Kagoshima University,  \\
1-21-35 Korimoto, Kagoshima 890-0065}
 \email{hiroimai@sci.kagoshima-u.ac.jp}

\altaffiltext{2}{International Centre for Radio Astronomy Research, M468, \\
The University of Western Australia, 35 Stirling Hwy, Crawley, Western Australia, 6009} 

\altaffiltext{3}{Center for Fundamental Education, Teikyo University of Science, \\
2525 Yatsusawa, Uenohara, Yamanashi 409-0193}

\altaffiltext{4}{Department of Physics, University of Hong Kong, 
Pokfulam Road, Hong Kong, China}

\altaffiltext{5}{Mizusawa VLBI Observatory, National Astronomical Observatory of Japan, \\
2-21-1 Osawa, Mitaka, Tokyo 181-8588}

\altaffiltext{6}{Nobeyama Radio Observatory, National Astronomical Observatory of Japan, \\
Minamimaki, Minamisaku, Nagano 384-1305}

\altaffiltext{7}{Korean VLBI Network, Korea Astronomy and Space Science Institute, \\
P.O. Box 88, Yonsei University, Seongsan-ro 262, Seodaemun, Seoul 120-749, Republic of Korea}


\begin{document}

\maketitle


\begin{abstract}
In this {\it Letter}, we report detections of SiO $v=3$ $J=1\rightarrow 0$ maser emission in very long baseline interferometric (VLBI) observations towards 4 out of 12 long-period variable stars: WX~Psc, R~Leo, W~Hya, and T~Cep. The detections towards WX~Psc and T~Cep are new ones. We also present successful astrometric observations of SiO $v=2$ and $v=$3 $J=1\rightarrow 0$ maser emissions associated with two stars: WX~Psc and W~Hya and their position-reference continuum sources: J010746.0$+$131205 and J135146.8$-$291218 with the VLBI Exploration of Radio Astrometry (VERA). The relative coordinates of the position-reference continuum source and SiO $v=3$ maser spots were measured with respect to those of an SiO $v=2$ maser spot adopted as fringe-phase reference. Thus the faint continuum sources were {\it inversely} phase-referenced to the bright maser sources. It implies possible registration of multiple SiO maser line maps onto a common coordinate system with 10 microarcsecond-level accuracy.
\end{abstract}


 \section{Introduction}

Silicon monoxide (SiO) maser emission has been used as an important probe of the dynamical structure and the physical condition of the inner circumstellar envelopes (CSEs) of asymptotic giant branch (AGB) and post-AGB stars. The pumping mechanism of the SiO masers is still an open question and understanding of it is essential to the diagnostics of the CSEs through observed behaviors of clump clusters of the masers such as temporal variations of the flux density, angular distribution, and three-dimensional velocity structure. SiO maser emissions of $v=1$ $J=1\rightarrow 0$, $v=2$ $J=1\rightarrow0$, and $v=1$ $J=2\rightarrow1$ have been main targets of very long baseline interferometric (VLBI) observations (e.g., \cite{sor04} and references therein). The $v=3$ $J=1\rightarrow 0$ maser line is also a unique target (\cite{ima10}, hereafter Paper {\rm I}; \cite{des12}). This transition is located at an energy level higher by $\sim$4 000~cm$^{-1}$ ($\sim$5 800~K) than the rotational transitions in the vibrational ground state and it needs considerably strong excitation in the gas at a temperature of 2 000--3 000~K of the surface of AGB and post-AGB stars. Observations of this maser line may be a good test for currently most plausible maser pumping model (line-overlapping, \cite{sor04} and references therein). However, its detection in VLBI observations is difficult due to its extreme weakness \citep{cho96,nak07}. 

By the way, precise measurement of relative positions of maser spots in the different maser transitions is essential for correctly deducing the pumping mechanism of SiO masers. Therefore,  the registration technique of multiple SiO maser line maps onto a common coordinate system is always an interesting issue and should be improved. In any technique, accurate determination of the absolute coordinates of maser spots and accurate evaluation of the errors associated with instrumental factors (e.g., accuracy of VLBI station coordinates on the terrestrial reference frame) are major factors to reduce the map registration uncertainty. Straightforward measurement of maser spot positions with respect to extragalactic continuum sources (e.g., \cite{zha12} and references therein) yields a 10 microarcsecond($\mu$as)-level accuracy of the map registration. Note that such a position-reference continuum source is often too faint to detect within a VLBI coherence time in the 40~GHz band ($\lesssim$2~min). Therefore, a special technique for phase-referencing coherent integration is required \citep{ima13}. 

In this {\it Letter}, we report new detections of SiO $v=3$ $J=1\rightarrow 0$ maser emission towards two long-period variable stars: WX~Piscium (WX~Psc) and T~Cephei (T~Cep) and two repetitive detections of the $v=3$ masers towards R~Leonis (R~Leo)(its first detection was reported by \cite{des12}) and W~Hydrae (W~Hya) in VLBI observations with the telescopes of Nobeyama Radio Observatory (NRO) and the VLBI Exploration of Radio Astrometry (VERA). The relatively high-sensitivity and short baselines including the NRO 45~m telescope enabled us to detect the faint $v=3$ masers. We also report successful phase-referencing VLBI observations of SiO $v=2$ and $v=3$ $J=1\rightarrow 0$ maser emissions associated with WX~Psc and W~Hya conducted with the VERA telescopes in the dual-beam mode. With VERA's baselines (1000--2300~km), only compact unresolved maser spots are detectable, so it is difficult to analyze and discuss the difference in the maser distributions between the different transitions of SiO masers. Instead, the VERA astrometry provides a good anchor of position reference for SiO map registration as mentioned above. This {\it Letter} focuses on technical points of VERA astrometry for yielding a 10~$\mu$as-level accuracy of map registration and pays attention to providing a guideline for such astrometric observations in the 40~GHz band in which continuum source flux densities are usually lower than those in lower frequency bands and below a detection limit of $\sim$300~mJy with VERA within VLBI coherence time. 

 \begin{table*}[t]
 \caption{Parameters of pairs of the SiO maser star and the position-reference source in VERA astrometry}
 \label{tab:sources}
\vspace{-4mm}
\begin{center}
\begin{tabular}{l@{ }l@{}c@{}r@{ }r@{ }c@{ }rr} \hline\hline
Pair & & Scan\footnotemark[a] 
& \multicolumn{2}{c}{\underline{\hspace*{8mm}Phase tracking center\hspace{8mm}}} & 
$V_{{\rm ref}, v=2}$\footnotemark[b] & Synthesized\footnotemark[c] & 1-$\sigma$ noise level\footnotemark[d] \\  
ID & Source name & (hr) & R.A (J2000) & Decl. (J2000) & (km~s$^{-1}$) & beam pattern & (mJy~beam$^{-1}$) \\
\hline
1\dotfill & WX~Psc & 2.6 & 01$^{\rm h}$06$^{\rm m}$25$^{\rm s}$\hspace{-2pt}.99 
& $+$12\arcdeg 35\arcmin 53\arcsec\hspace{-2pt}.4 & 7.4 & 0.38$\times$0.69, $-$39\arcdeg & 194, 201 \\
& J010746.0$+$131205 & & 01$^{\rm h}$07$^{\rm m}$45$^{\rm s}$\hspace{-2pt}.961872
& $+$13\arcdeg 12\arcmin 05\arcsec\hspace{-2pt}.19067 & ... & 0.47$\times$0.64, $-$48\arcdeg & 3.9\\
\hline
2 \dotfill& W~Hya & 2.8 & 13$^{\rm h}$49$^{\rm m}$01$^{\rm s}$\hspace{-2pt}.9311 
& $-$28\arcdeg 22\arcmin 04\arcsec\hspace{-2pt}.560 & 40.4 & 0.44$\times$1.17, $-$19\arcdeg & 184, 184 \\
& J135146.8$-$291218 & & 13$^{\rm h}$51$^{\rm m}$46$^{\rm s}$\hspace{-2pt}.838765
& $-$29\arcdeg 12\arcmin 17\arcsec\hspace{-2pt}.65002 & ... & 0.43$\times$1.15, $-$18\arcdeg & 10.6 \\
\hline
\end{tabular}
\end{center}
 \noindent
 \footnotemark[a]Total integration time. \\
 \footnotemark[b]LSR velocity of the SiO $v=2$ $J=1\rightarrow 0$ maser spot adopted as phase-reference. \\
 \footnotemark[c]Major and minor axis length in unit of mas and position angle. \\
 \footnotemark[d]In maser image cubes, emission-free spectral channel maps of the $v=2$ and $v=3$ lines were chosen for the calculation. \\
 \end{table*}


\section{Observations and data reduction}

We conducted VLBI observations of SiO $v=2$ and $v=3$ $J=1\rightarrow$0 maser emissions (at rest frequencies of 42.820582 and 42.519340~GHz, respectively) on 2012 March 24--25 and 2012 May 20--21 for 25~hr each towards 12 stars in total: WX~Psc, AP~Lyn, U~Ori, VY~CMa, R~Leo, RS~Vir, W~Hya, U~Her, RU~Her, V1111~Oph, V4120~Sgr, and T~Cep. They were selected from the sources detected in the SiO $v=3$ $J=1\rightarrow 0$ maser emission by \citet{cho96}. Three or four VERA$^{1}$ 20~m telescopes and the 45~m NRO\footnote
{The NRO and VERA/Mizusawa VLBI observatory are branches of the National Astronomical Observatory of Japan, an interuniversity research institute operated by the Ministry of Education, Culture, Sports, Science and Technology.} 
were simultaneously operated; the former for observing the target maser and position-reference continuum sources using the dual-beam receiving system of VERA and the latter for the maser sources only but with higher sensitivity. Both of the VERA and NRO telescopes also observed bright continuum calibrators every 40~min for calibration of instrumental group-delay and fringe-phase residuals and bandpass characteristics. The observed signals, received in left-hand circular polarization, were digitized in four levels and divided into 16 base band channels (BBCs) each with a band width of 16 MHz. In the VERA stations, they were recorded with both the SONY DIR1000 and DIR2000 recorders at rates of 128 and 1024~M~bits~s$^{-1}$, respectively. Two of the BBCs were assigned to the SiO $v=2$ and $v=3$ masers in one beam and others to the position-reference continuum source emission in other beam. The DIR1000 recording was also made in NRO and accepted only the two SiO maser BBCs. The data correlation was processed with the Mitaka FX correlator, in which each of the maser BBCs was split into 512 spectral channels, corresponding to a velocity spacing of 0.22\kms. On the other hand, each of the continuum BBCs was split into 32 spectral channels. 

In March, each of maser--continuum source pairs was observed for only 2--3 hr. Because the DIR1000 and DIR2000 data were available from only three and four stations, respectively, the imaging and the astrometry of maser sources were difficult. After the normal calibration procedures as mentioned soon later, we obtained the SiO $v=2$ and $v=3$ maser spectra with high sensitivity by coherently integrating the whole visibility data. In May, the observations similar to those in March were conducted, but the maser sources with the $v=3$ detections in March had longer scans (for 4--5~hr) for more surely successful maser source astrometry with VERA. In this {\it Letter}, we focus on the spectra of SiO $v=2$ and $v=3$ $J=1\rightarrow 0$ maser emissions towards WX~Psc, R~Leo, W~Hya, and T~Cep with the successful $v=3$ detections in the NRO--VERA baselines in March and the successful astrometric results of WX~Psc and W~Hya in the VERA dual-beam observations in May. The results of the whole observations including all SiO $v=2$ and $v=3$ maser maps will be published in a separate paper. 

Data reduction and image synthesis were made using the NRAO AIPS package. In order to conduct the astrometry using faint continuum sources, the {\it inverse} phase-referencing technique was adopted, which is described in more detail in \citet{ima13}\footnote
{We used the ParselTongue/python pipeline scripts through the data reduction, which are now available in the wiki page: 
http://milkyway.sci.kagoshima-u.ac.jp/groups/vcon\_lib/wiki/9fbfd/Data\_Analysis.html .
See also the ParselTongue wiki page: http://www.jive.nl/dokuwiki/doku.php?id=parseltongue: parseltongue .}.
First, calibration of visibility amplitudes for antenna gains and bandpass characteristics and that of visibility phases for instrumental group-delay and phase residuals were made in a standard manner by using scans on the continuum calibrators. Then, fringe fitting and self-calibration were performed using visibilities in the velocity channel (Column 6 of Table \ref{tab:sources}), which includes a $v=2$ bright maser spot (velocity component). The Doppler velocity is given with respect to the local standard of rest (LSR). The solutions were applied to the data in other velocity channels of the $v=2$ maser and those of the $v=3$ maser as well as to the data of the position-reference source. In the May session, thus we obtained the maser image cubes and the reference source maps with naturally weighted visibilities, which yielded a typical size of the synthesized beam of 0.5 milliarcsecond (mas). 

 \begin{table}[t]
 \caption{Results of the VERA astrometry}
 \label{tab:astrometry}
 \vspace{-2mm}
\begin{center}
\begin{tabular}{l@{}c} \hline\hline
Source name & Offsets (mas)\footnotemark[a] \\  
\hline
WX~Psc\dotfill & ($-24.24\pm0.05$, $-453.14\pm0.03$) \\
W~Hya\dotfill & ($240.00\pm0.01$, $263.67\pm0.03$) \\
\hline
\end{tabular}
\end{center}
 \noindent
 \footnotemark[a]Measured (R.A., decl.) offsets with respect to the phase-tracking center shown in 
 Table \ref{tab:sources}. \\ 
 \end{table}


 \begin{figure*}[p]
 \begin{center}
     \FigureFile(160mm,50mm){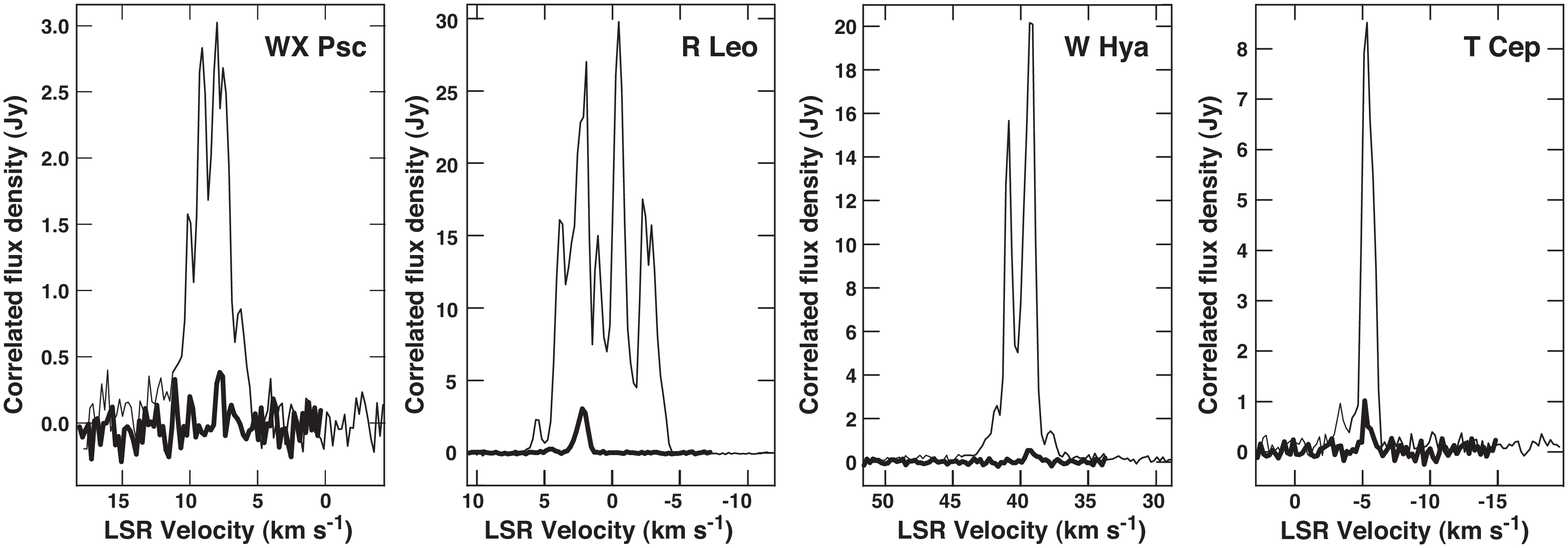}
 \end{center}
    \vspace{-4mm}     
    \caption{SiO $v=2$ (thin line) and $v=3$ (thick line) $J=1\rightarrow 0$ maser spectra obtained towards WX~Psc, R~Leo, W~Hya, and T~Cep on 2012 March 24--25. The visibility data were scalar-averaged and integrated with the DIR1000 data including the 45~m telescope. }
 \label{fig:spectra}

 \begin{center}
     \FigureFile(150mm,50mm){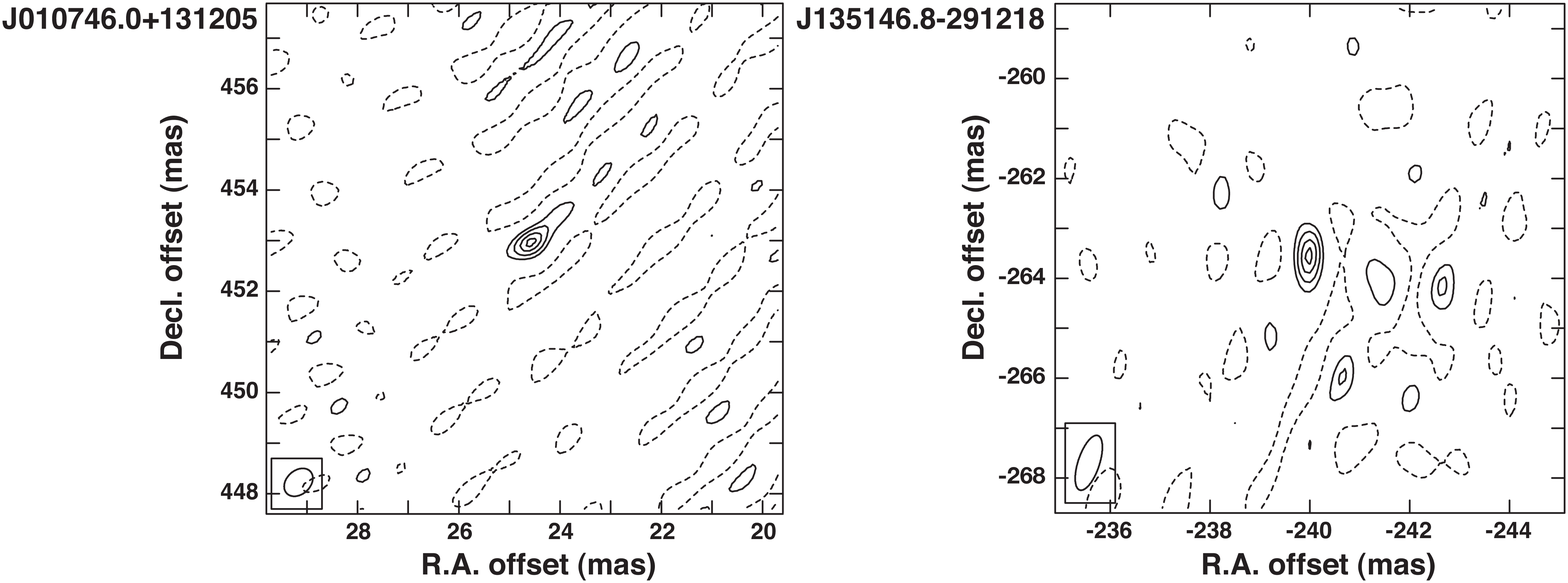}
 \end{center}
 \vspace{-5mm}
    \caption{Contour maps of the position-reference continuum sources: J010746.0$+$131205 (left panel) and J135146.8$-$291218 (right panel). The contour levels are set at $-$26, 52, 103 and 122 mJy~beam$^{-1}$ in the left panel and at $-$6, 11, 17, 23, and 27 mJy~beam$^{-1}$ in the right panel. The synthesized beam pattern is displayed in the bottom-left corner box.}
 \label{fig:continuums}
 
     \FigureFile(175mm,80mm){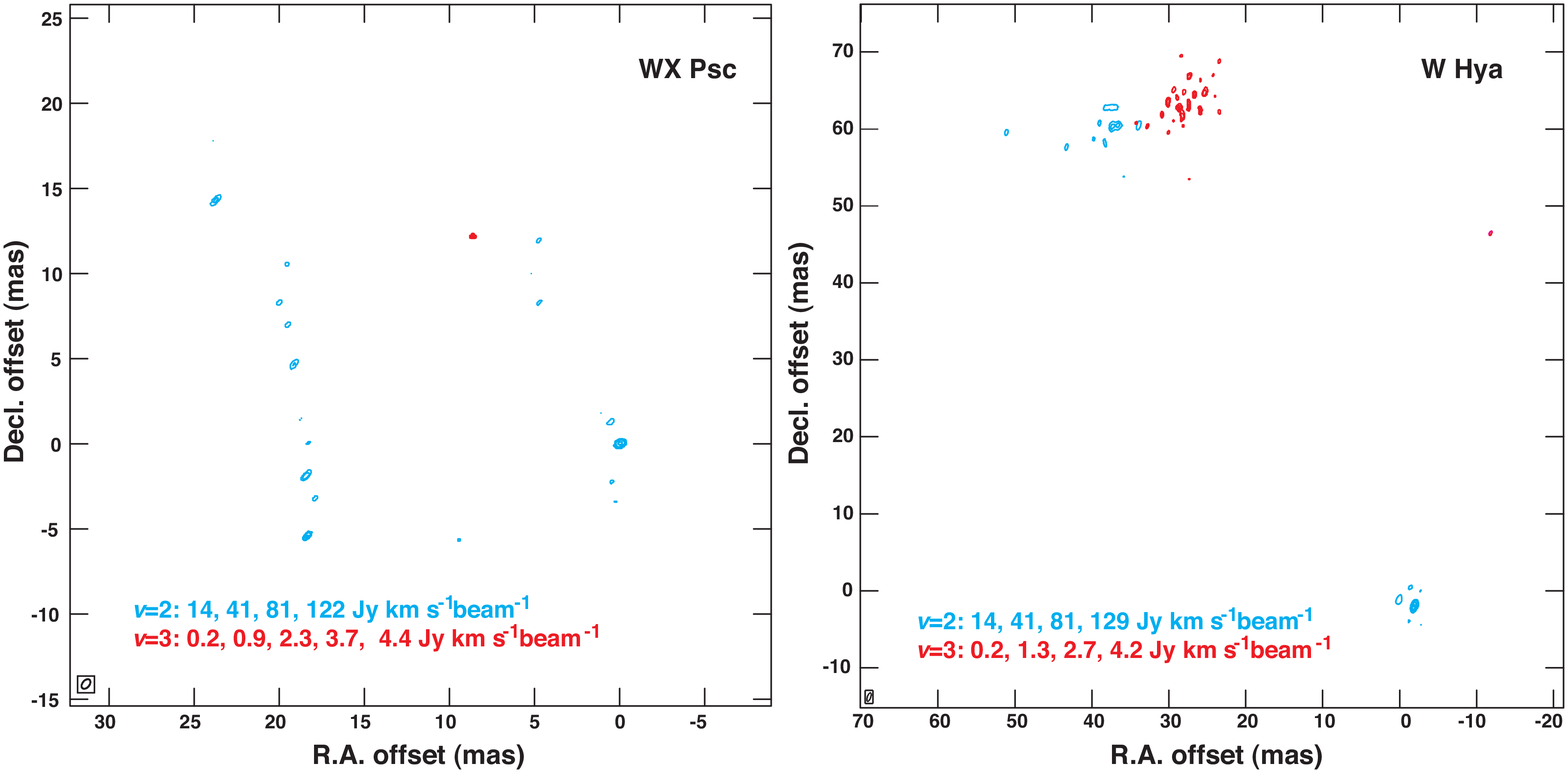}
    \vspace{-2mm}
    \caption{Velocity-integrated brightness contour maps of SiO $v=$2 (cyan) and $v=$3 (red) $J=1\rightarrow 0$ maser emissions associated with WX~Psc (left panel) and W~Hya (right panel). The contour levels are shown in the panel. The synthesized  beam pattern is displayed in the bottom-left corner box. Noise pixels below a 6-$\sigma$ level were blanked before synthesizing these maps.}
 \label{fig:masers}
\end{figure*}


 \section{Results and Discussion}
 \label{sec:results}

Figure \ref{fig:spectra} shows the SiO $v=2$ and $v=3$ $J=1\rightarrow 0$ maser spectra with the successful $v=3$ detections. All of these maser sources were observed in the nearly stellar light maxima (during the light curve phase of $\phi\approx$0.8--1.0). The $v=3$ maser was again detected toward W~Hya (with a pulsation period of $P\approx$360~d) three years after the first VLBI detection (Paper I). They are consistent with the suggestion that the $v=3$ maser should have strong correlation with the stellar light curve \citep{cho07}. The $v=3$ masers seem to be associated with relatively bright $v=2$ maser components on the spectra, however as mentioned later, they are not necessarily spatially coincident with each other. 

Table \ref{tab:sources} gives parameters of the individual pairs of maser and position-reference sources with the successful astrometric results. It includes the parameters of the observation, signal correlation, and data reduction. The counterpart continuum sources of these maser sources, J010746.0$+$131205 and J135146.8$-$291218, respectively, are separated by 1\arcdeg.03 and 0\arcdeg.69 from the masers. Figure \ref{fig:continuums} shows the phase-referenced images of the two position-reference sources. The 1-$\sigma$ noise levels of the images are 3.4 and 10.1~mJy~beam$^{-1}$, respectively. The measured position offset of the continuum source (in Figure \ref{fig:continuums}) is inversely equal to that of the phase-reference $v=2$ maser spot from the phase-tracking center of the maser source data (Columns 4 and 5 in Table \ref{tab:sources}). Table \ref{tab:astrometry} gives the determined coordinate offsets of the phase-reference $v=2$ maser spots. Because the absolute coordinates of the reference sources are determined with an accuracy better than 0.5~mas (e.g. \cite{pet08})\footnote{See also URL: http://astrogeo.org/vlbi/solutions/rfc\_2012b .}, 
those of the reference maser spots are also determined in this level of accuracy. The position drift of the $v=3$ maser map with respect to the $v=2$ map appears proportionally to the position offset of the phase-reference maser spot and the frequency difference between the $v=2$ and $v=3$ masers (\cite{gwi92}; Paper I). Using the derived position offsets mentioned above, this position drift was corrected for the $v=3$ maps.\footnote
{Phase correction using the AIPS task CLCOR before fringe-fitting with calibrator scans can perform the equivalent position correction.} In the present VLBI observations, the map registration of the $v=2$ and $v=3$ maser lines has an uncertainty less than 50~$\mu$as,  small enough to discuss the maser pumping models.

Figure \ref{fig:masers} shows the composite maps of SiO $v=2$ and $v=3$ $J=1\rightarrow 0$ maser lines in WX~Psc and W~Hya. 
The maser emissions were significantly spatially resolved; the cross-power flux densities of the $v=2$ and $v=3$ masers were much lower than the total-power flux densities ($\sim$400 and $\sim$50~Jy in WX~Psc and $\sim$1000 and $\sim$350~Jy in W~Hya, respectively). Comparing with those of previous observations (e.g., \cite{sor04} and Paper {\rm I}, respectively), only for unresolved, compact maser spots, we find that their relative positions have roughly persisted for a decade. Assuming positions of the central stars at ($\sim$13, $\sim$5) and ($\sim$25, $\sim$30) [mas] for WX~Psc and W~Hya, respectively, the distance to the $v=3$ spots from the star is roughly equal to that of the $v=2$ spots, consistent with the previous observations (Paper {\rm I}; \cite{des12}). As already mentioned, the $v=3$ and $v=2$ maser spots were not spatially coincident, in contrast to the results of Paper I and \citet{des12}. These results favor a collisional pumping scheme for SiO masers (e.g., \cite{loc92,hum02}). However, it is premature to draw ring-shaped distribution models in Figure \ref{fig:masers} and to discuss the correlation of the $v=2$ and $v=3$ emission distributions in the present results in which other fainter and more extended maser spots are missing. Thus we cannot rule out a scheme of line-overlapping between SiO and \h2o\ ro-vibrational transitions,  not only for $v=2$ $J=1\rightarrow 0$ \citep{sor04} but also $v=3$ $J=1\rightarrow 0$. \citet{cho07} consider the most plausible overlap pair between the SiO $v=2$ $J=0 \rightarrow$ $v=3$ $J=1$ and the \h2o\ $\nu_{2}=2$ $5_{05} \rightarrow$ $\nu_2=1$ $6_{34}$ lines. This overlapping can excite the SiO $v=3$ $J=1\rightarrow$0 maser. Temporal variation in spot displacement of different maser lines as predicted by \citet{hum02} can be examined by future monitoring observations with combination of high precision astrometry as those demonstrated in this paper and high dynamic imaging of the maser sources (e.g., \cite{yi05}). 

Here we emphasize that the present success in long-time coherent integration using the {\it inverse} phase-referencing technique is a milestone for VERA astrometry using maser--continuum source pairs in the 40~GHz band. In the maser source astrometry with the Very Long Baseline Array (VLBA)(e.g. \cite{zha12}), SiO maser sources have been often used as phase-reference because of their detections at signal-to-noise ratios higher than those of continuum sources. On the other hand, taking into account more complicated procedures of the inverse phase-referencing technique \citep{ima13}, a relatively high signal recording rate of VERA has favored continuum sources as phase-reference. Therefore, successful astrometric observations had been limited to nearly ideal cases in which both of maser and continuum sources should be bright enough to detect within a VLBI coherence time (e.g., \cite{cho09}). Such a condition empirically demands flux densities higher than $\sim$20~Jy and 300~mJy for SiO maser and 40~GHz continuum sources for VERA astrometry, respectively. The present results prove that a 30~mJy-level continuum source is detectable with the {\it inverse} phase-referencing technique by using data of a bright SiO maser source. This has advantage over the continuum-reference case in which even a bright, well-studied SiO maser source was dropped out from the target list of VERA astrometry because of the absence of bright continuum sources within 0.5\arcdeg--2.2\arcdeg\ of the maser source for the VERA dual-beam observations. As a result of the milestone achievement, a statistically moderate number ($>$10) of well-studied SiO maser sources harboring multiple transitions of SiO masers, including $v=3$ $J=1\rightarrow 0$, will become targets for the study on pumping mechanism of SiO masers in VLBI and for the trigonometry with VERA.  Moreover, in such a situation, we have to concern only about the dispersion of maser spot distributions, which is much larger than the uncertainty of relative positions among the different lines of SiO masers. This is helpful for unambiguous interpretation of the pumping mechanism of the SiO masers.

\bigskip
We acknowledge all staff members and students who have helped in array operation and in data correlation of VERA. 
HI has been supported for stay at ICRAR by the Strategic Young Researcher Overseas Visits Program for Accelerating Brain Circulation funded by Japan Society for the Promotion of Science (JSPS).
JN has been supported by the Research Grants Council of Hong Kong (project code: HKU 703308P, HKU 704209P, HKU 704710P, and HKU 704411P), and the Small Project Funding of the University of Hong Kong (201007176004).

\end{document}